# Fundamental Theory of the Evolution Force: Gene Engineering Utilizing Synthetic Evolution Artificial Intelligence

L. K. Davis

*Abstract*—The effects of the evolution force are observable in nature at all structural levels ranging from small molecular systems to conversely enormous biospheric systems. However, the evolution force and work associated with formation of biological structures has yet to be described mathematically or theoretically. In addressing the conundrum, we consider evolution from a unique perspective and in doing so introduce the "Fundamental Theory of the Evolution Force: *FTEF*". Herein, we prove *FTEF* by proof of concept using a synthetic evolution artificial intelligence to engineer 14-3-3 ζ docking proteins. Synthetic genes were engineered by transforming 14-3-3 ζ sequences into time-based DNA codes that served as templates for random DNA hybridizations and genetic assembly. Application of time-based DNA codes allowed us to fast forward evolution, while damping the effect of point mutations. Notably, SYN-AI engineered a set of three architecturally conserved docking proteins that retained motion and vibrational dynamics of native *Bos taurus* 14-3-3 ζ.

## I. Introduction

THE evolution force may be described as a compulsion acting at the matter-energy interface that drives molecular diversity while simultaneously promoting conservation of structure and function. The effects of the evolution force are manifested at all levels of life and are responsible for such processes as formation of genes and gene networks. Herein, we introduce the "Fundamental Theory of the Evolution Force (*FTEF*)" and utilize the *FTEF* to predict the formation of genomic building blocks (GBBs). From our perspective GBBs are short highly conserved sequences formed as evolution artifacts and are principal components of genes. It is not difficult to assert that DNA and protein are matter based computer programs. When viewing genes from the perspective of a computer algorithm GBBs are analogous to fundamental programming blocks. In the current study, we designed a synthetic evolution artificial intelligence (SYN-AI) to identify evolution force promoting formation of these programming blocks and to engineer genes by their assembly.

The *FTEF* is based on four evolution force identifiers, (i) evolution conservation, (ii) wobble, (iii) DNA binding state, and (iv) periodicity that allow comparison of the magnitude of evolution force associated with DNA crossovers and genomic building block formation. While a strong association between cellular function and evolutionary conservation of DNA and protein sequence has long been recognized [1]-[5], wobble is classically defined as genetic diversity within the third codon with conservation of amino acid sequence [6]-[12]. Herein, we expand wobble's definition to encompass the achievement of genetic diversity with simultaneous conservation of structure, thusly allowing wobble to be quantifiable at all structural levels. We establish DNA binding states as evolution force identifiers based on the assumption that the association of energy and life is inseparable, and we assert that interaction of the evolution force at the matter-energy interface may be characterized by DNA binding states [13]-[16]. Additionally, there exists strong correlations between sequence periodicity and conservation of structure and function as described in [17]-[19]. Thusly, we propose that periodicity is an indicator of evolution force. Prominently, we show that application of these four identifiers in conjunction with selection pressure is sufficient to engineer genes de novo.

In order to simulate evolution, SYN-AI integrates a gene-partitioning model that assumes contemporary genes evolved from a single ancestor that expanded to the modern gene pool. Thusly, *FTEF* is in agreement with the "Universal Ancestor" and LUCA "Last Universal Common Ancestor" models, [20], [21]. We reconstruct DNA exchanges occurring during gene evolution and subsequent point mutations due to speciation by performing gene partitioning. Gene sequences are transformed into DNA secondary (DSEC) and tertiary (DTER) codes in correlation with protein hierarchical structure levels. Thusly, we introduce an arbitrary time dimension to the DNA code that allows us to fast-forward evolution while dampening the effects of point mutations that lead to disruption of protein structure. The application of time-based DNA codes allows for conservation of both global and local protein architecture as genomic building blocks are conserved from LUCA and have been tested by the evolution process. In terms of hierarchical structure, the DSEC simulates evolution on the genomic building block scale in the range of 19 – 21 base pairs, wherein the DTER simulates evolution at the super secondary structure level based on protein quaternary structure. Thusly, the exchange of genetic information during synthetic evolution is synonymous to the swapping of GBBs in a game of Legos and agrees with the 'Domain Lego' principle [22], [23].

We proved *FTEF* by proof of concept employing SYN-AI to engineer a set of 14-3-3 ζ docking proteins using the *Bos taurus* 14-3-3 ζ docking gene as an engineering template.

L.K. Davis was with the Cooperative Agriculture Research Center (CARC) at Prairie View A & M University, Prairie View, TX 77446 USA. He is now at the Gene Evolution Project, LLC (e-mail: lkdavis.geneevolutionproject@gmail.com).

Genomic building blocks were identified by the magnitude of evolution force associated with DNA crossovers. Whereby, DNA crossovers were simulated by random hybridization of DNA fragments within genomic alphabets comprising the DNA secondary code. Synthetic super secondary structures were engineered based upon the DNA tertiary code and were constructed by random selection and ligation of genomic building blocks. Following equilibration of synthetic structure lengths to native structures, we simulated natural selection by applying Blosum 80 and PSIPRED secondary structure-based algorithms to select synthetic super secondary structures for gene engineering. Synthetic docking genes were engineered by randomly selecting and ligating synthetic structures from appropriate DTER libraries. Notably, SYN-AI constructed a library of 10 million genes that yielded three architecturally conserved 14-3-3 ζ docking proteins based on the theoretical closeness of their hydrophobic interfaces and active sites to the native *Bos taurus* docking protein.

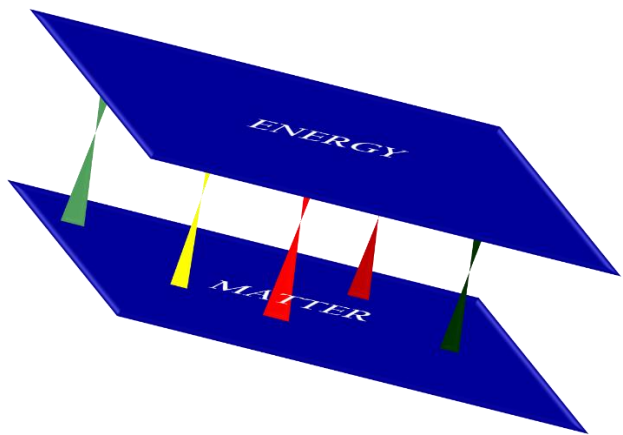

Fig. 1 The Matter-Energy Interface

## II. THEORY

### A. FTEF

We state herein that the evolution force is a compulsion acting at the matter-energy interface that drives genetic diversity while simultaneously conserving biological structure and that the dynamics of the matter-energy interface do not act independently of evolution's tendency toward conservation. We further hypothesize that the four principal identifiers of evolution force are 1) evolution conservation, 2) wobble, 3) DNA binding state, and 4) periodicity.

We established these evolution force identifiers according to the basic engineering format that nature utilizes in respect to genetic relatedness as well as established evolution concepts. To give a simple explanation of *FTEF*, when considering evolutional conservation of structure, we can use the example of bone structure. Human legs comprise of an upper leg having a femur and a lower leg comprising of a tibia and fibula. These structures are conserved in a variety of species in the phylum Chordata, thusly we consider them as artifacts of the evolution force. When determining wobble, *FTEF* views these conserved structures with respect to the range of genetic diversity covered as they are conserved in genetically distant species. In terms of their periodicity, the *FTEF* hypothesizes that the more frequently such structures are observed in nature the stronger the influence of the evolution force.

We utilize the *FTEF* to describe these evolution principals at the molecular level and to engineer genes. However, our theory may be applied to all levels of life. While, the concepts of evolution conservation, wobble and periodicity are straight forward, our conceptualization of a matter-energy interface requires more clarification due to the Theory of Quantum Mechanics and the coexistence of photons as both particles and waves. Not all energy manifest as matter but also in various forms of kinetic and thermal energy, thusly to describe the effect of energy on gene evolution we took an alternative approach. The *FTEF* views energy and matter as separate but overlapping dimensions that form synapse at critical junctions allowing the sharing of information, Fig. 1. These interfaces are often observed in nature such as the interface of sound waves with the ocular allowing transduction of vibrational energy and its conversion to information by the brain. More ubiquitously, photons interact with photoreceptors allowing for the conversion of radiation to cellular information and in plants its conversion to chemical energy as glucose. In terms of gene evolution, DNA crossover junctions are a type of matter-energy interface that allow for conversion of thermal energy to genetic information. Where, enthalpic and entropic factors such as divalent cation concentrations and temperature governed by cellular conditions and sequence contribute to stabilization of the DNA molecule and facilitate the transfer of genetic information.

The *FTEF* states that evolution force associated with formation of genomic building blocks may be solved for according to the postulates stated below:

*Postulate 1* - A natural selection system will generate sequences exhibiting positive variation from the mean of a population of randomly evolved sequences occurring during an evolution instance. Whereby, such sequences will display greater evolutionary conservation of the parental sequence.

*Postulate 2* - Due to degeneracy of the genetic code [8], a natural selection system will generate sequences that exhibit higher conservation of protein structure than expected based on mean DNA similarity. This tendency is defined as wobble and considered an artifact of the evolution force.

*Postulate 3* - Evolution force regulates molecular diversity at the matter-energy interface in the form of Gibb's free energy dependent DNA base stacking interactions. Thusly, evolution force may be characterized by DNA binding states.

*Postulate 4* - Evolution tends to repeat structures that contribute to survival, whereby structures that contribute to function occur more frequently. Thusly, evolution force may be solved as a function of sequence periodicity.

### B. Evolution Force Identifiers

1) Evolutionary Sequence Conservation

Sequence conservation is strongly correlated with residues associated with ligand binding and active sites, protein-protein interaction (PPI) and functional specificity [1]. In a study of

DNA binding proteins, it was shown that functionally essential residues are more highly conserved than their counterparts [24] and are associated with tightly packed sites that play a role in the protein stable core or indicative of folding nucleus [25]. Relatedly, it has been shown that genes that encode proteins involved in numerous protein-protein interactions such as 14-3-3 ζ docking proteins are more evolutionarily conserved than genes encoding less-prolific interactors [26].

Our theory agrees with the Fundamental Theory of Natural Selection as it captures the effects of fitness on gene evolution by identifying genomic building blocks that characterize fit haplotypes [27]. The *FTEF* does so by considering all possible DNA crossovers in an evolution instance. Note, that the term DNA crossovers refers to hybridizations of genomic building blocks in a SYN-AI cycle. These short gene sequence blocks form by numerous DNA exchanges occurring over evolution of orthologue/paralogue sequence space and encode diverse functions due to speciation. Thusly, the *FTEF* simulates time-development of gene sequence space and engineers genes by the assemblage of highly evolved sequence blocks.

GBBs are identified based on the magnitude of evolution force applied about evolution conservation engine $\epsilon$, where $\epsilon$ describes conservation at DNA and protein levels and is a function of evolution vectors $\epsilon_{DNA}^c$ and $\epsilon_{Pro}^c$. These position vectors characterize DNA crossover homology to the parent sequence in respect to a rigid body of sequences that comprises full enumeration of DNA crossovers occurring over an evolution instance. They report the position of DNA and protein sequences resulting from DNA crossovers in the evolution potential field and are functions of similarity vectors $X_i$ and $X_j$ that compare recombinant DNA and protein sequences to parental in terms of physiochemical properties, volume, hydrophobicity and charge. The rigid body generates the evolution potential field, wherein relative position of DNA crossovers describes their evolutional advantageousness with more distant DNA crossovers being more evolutionarily advantageous. Relative positions are described by weighting similarity vectors by evolutional weights $W_d$ and $W_p$ as given in (2) and (3). By applying these weights, we normalize the relative position of a sequence in the configuration space to all other DNA crossovers and characterize the full enumeration of DNA crossovers back to LUCA. Whereby, the configuration space describes the evolutionary history of the gene.

$$\epsilon = \epsilon_{DNA}^c \cdot \epsilon_{Pro}^c \quad (1)$$

$$\epsilon_{DNA}^c = W_d \sum_{i=1}^{GBB} X_i, \quad i = nucleotide \quad (2)$$

$$\epsilon_{Pro}^c = W_p \sum_{j=1}^{GBB} X_j, \quad j = residue \quad (3)$$

Evolution weights $W_d$ and $W_p$ describe the rigid body's center of gravity, thusly describe the origin of the evolution potential field. They are functions of recombinant pool mean DNA $\mu_s^{DNA}$ and protein $\mu_s^{Prot}$ similarity vectors, thusly describe positions of all DNA crossovers in the potential field. They are solved by the summation of DNA $X_i$ and protein $X_j$ similarity vectors occurring within sequence space ($sspace^r$). Where, $sspace^r$ comprises all orthologue-paralogue gene sequences at a selected identity threshold. Evolutional weight is solved in respect to the total number of DNA crossovers (N), thusly reflects full enumeration of DNA crossovers occurring within the evolution potential field.

$$W_d = \frac{1}{n\mu_s^{DNA}} \text{ and } \mu_s^{DNA} = \frac{1}{N}\left[\sum_{DNA=1}^{sspace^r} \sum_{i=1}^{GBB} X_i/n\right] \quad (4)$$

$$W_p = \frac{1}{n\mu_s^{Pro}} \text{ and the } \mu_s^{Pro} = \frac{1}{N}\left[\sum_{Prot=1}^{sspace^r} \sum_{j=1}^{GBB} X_j/n\right] \quad (5)$$

### 2) Molecular Wobble

Wobble evolved during expansion of the genetic code from a simple triplet code expressing a few amino acids in which only the middle position was read as proposed by Crick [28] to the modern genetic code comprising 64 codons and 20 amino acids. This is corroborated by Wu, whom suggested evolution of the modern code from an intermediate doublet system, where only the first and second codon positions were read and the third position served as a structural stabilizer [29]. These hypotheses are substantiated by evolution remnants found in aminoacyl tRNA synthetases that support evolution of the modern genetic code from a more primitive ancestor [30]. Moreover, they support the "*Coevolution Theory*" which suggests the genetic code is an imprint of prebiotic pathways that evolved over a three-billion-year period and that were fixed in LUCA [31].

Due to coevolution of wobble with the genome, *FTEF* views wobble as one of the four principal evolution force identifiers. Prominently, wobble allows the evolution force to balance fitness and adaptation by conserving protein sequence, while simultaneously introducing genetic diversity in the third codon position. Due to structure and grouping effects in the genetic code, mutations in neighboring codon positions also result in genetically close amino acids. Thusly, we define wobble in a more generic fashion allowing us to capture the property in all three codon positions. *FTEF* solves for wobble $\omega_m$ characterizing a DNA crossover by overlapping position vectors $\epsilon_{DNA}^c$ and $\epsilon_{Pro}^c$ (6), thusly does not discriminate the 3rd codon position. The resulting relationship is a good indicator of evolution force as it is reflective of parallel hierarchical sequence transitions defining multiple molecular states. *FTEF* designates wobble as a function of genetic displacement $x$ over time $t$, where $t$ is the number of evolution cycles required to achieve a genetic step of distance $x$. Displacement of the protein position vector respective to the DNA position vector in the configuration space is described by genetic step $x = (\epsilon_{Pro}^c/\epsilon_{DNA}^c) - i_n$. Where, $i_n$ is an element of identity vector $\hat{i}$ and characterizes expected positions of DNA crossovers in the evolution potential field. Expected positions characterize the mean of the recombinant pool and are defined as unit vectors, where $\forall i_n = 1$. Recombinations characterized by sequences displaying greater conservation of protein sequence than DNA display wobble.

$$\omega_m = \frac{x}{t}, \text{where } x = \frac{\epsilon_{Pro}^C}{\epsilon_{DNA}^C} - i_n, \ i_n \in \{\vec{\iota}\} \text{ and } \forall \ i_n = 1 \quad (6)$$

3) DNA Binding States

*FTEF* assumes that synthetic evolution processes simulated by SYN-AI mimic evolution, thusly DNA binding states occurring during simulations are analogous to DNA crossovers occurring during meiosis as supported by previous studies discussing the anticipatory effects of DNA shuffling [32]. Genetic diversity occurs by processes such as DNA crossovers and translocations that result in gene duplication, inversion, insertion and deletion [33], [34]. It is widely accepted that these processes result in relaxation of evolutionary stringency allowing speciation and random point mutations by neutral evolution [35], [36], whereby artifacts of these processes are captured in GBBs.

According to the *FTEF*, DNA crossover junctions are a matter-energy interface by which the evolution force conveys information. Thusly, the evolution engines introduced herein derive from and are dependent upon DNA binding states. The effect of the relationship between evolution conservation and sequence homology on DNA hybridization and Gibb's free energy is obvious. However, less obviously GBB frequency is also directly affected by DNA binding states occurring during gene duplication. Likewise, wobble evolved by convergence of environmental conditions on natural selection and ensuing speciation following DNA exchanges driven by DNA binding states. The inclusion of DNA binding states as an evolution engine allows *FTEF* to agree with complex theories describing the coevolution of genes and gene networks [37]. Whereby, formation of coevolution mechanisms described in Jordan is a consequence of DNA binding states that helped form genomic structural constraints, gene regulatory regions and nodes [37].

DNA binding states express the stoichiometric relationship between DNA crossovers, thusly account for thermodynamic contributors described by Gibb's free energy using less costly calculations. Thereby, we can track interaction of the evolution force at the matter-energy interface back to LUCA with less computational cost. DNA binding states $p^i$ are a function of annealing probability $A_{L-v,\ L}^V$ and DNA binding probability $P_{Keq}^i$ [14] and [15]. Thusly, they are a function of volume exclusion at the DNA crossover junction and DNA crossover thermodynamic signatures (7).

$$P^i = A_{L-v,L}^V \cdot P_{keq}^i \quad (7)$$

According to Wetmur, annealing probability $A_{L-v,L}^V$ distributes volume exclusion $V^\alpha$ [38] characterizing a DNA hybridization over that of the recombinant pool, where $V$ defines overlap length characterizing a DNA crossover and $L$ defines sequence length. Volume exclusion is a function of the length to volume relationship occurring at the DNA crossover junction, whereby the probability of hybridization decreases beyond a critical volume of the hybridization bubble.

$$A_{L-v,L}^V = d_v V^\alpha / \sum d_v V^\alpha, \text{where } \alpha = -\frac{1}{2} \quad (8)$$

Thermodynamic contributions to DNA binding states are described by equilibrium constant $k_{eq}$. Where, $k_{eq}$ of a DNA crossover is an exponential function of Gibb's free energy $\Delta G$. Gibb's free energy of hybridization is solved by summation of standard $G°(i)$ free energies for the 10 possible Watson-Crick nearest neighbors, whereby an entropic penalty $G°(sym)$ is incorporated for maintaining C2 symmetry [39]. Counterion condensation is accounted for by free energies of initiation $G°(init\ w/term\ G \cdot C)$ and $G°(init\ w/term\ A \cdot T)$.

$$P_{keq}^i = k_{eq} / \sum k_{eq} \quad (9)$$

$$k_{eq} = exp^{-\frac{\Delta G}{RT}} \quad (9a)$$

$$\Rightarrow exp^{-\frac{\sum_i n_i G°(i)\ +\ G°(init\ w/term\ G \cdot C)\ +\ G°(init\ w/term\ A \cdot T)\ +\ G°(sym)}{RT}}$$

4) Periodicity

Saliently, three base periodicity allows characterization of species based upon their Fourier spectrum [40]. Whereby, a strong peak at frequency 1/3 is observed in Fourier spectrums of genome coding regions suggesting the presence of selection pressure [41]. GBB periodicity results from gene duplication and subsequent speciation, wherein fit sequence blocks are retained by the genome. Thusly, building block periodicity reflects natural selection due to evolutionary fitness. The *FTEF* considers periodicity $P^\pi$ as an evolution force identifier and characterizes it as the distribution of GBB frequency $f_{ij}$ over its global frequency Z. Where, $f_{ij}$ describes oligonucleotide $i$ and peptide $j$ homolog occurrences within the target gene, and Z is a summation of occurrences within orthologue/paralogue sequence space. $P^\pi$ compares selectivity of a DNA crossover to adjacent sequences at both the DNA and protein level. Whereby, sequences displaying high periodicity are reflective of selection pressure by the evolution force.

$$P^\pi = \sum_i^{oligo} \sum_j^{peptide} f_{ij}^{gene} / Z \quad (10)$$
$$\text{where } Z = \sum_{n=1}^{sspace} \sum_i^{oligo} \sum_j^{peptide} f_{ij}$$

*C. Analyzing Evolution Force Utilizing the Linear Model*

The Linear Model (LM) considers evolution force both at the DNA and protein level and ignores transitory effects on mRNA transcripts. GBBs are viewed as particles having high momentum through an evolution potential field produced by a rigid body comprised of all DNA crossovers back to LUCA. We apply Newton's second law of motion to describe particle momentum $p = mv \Rightarrow \epsilon \cdot \omega_m$. Position vector $\epsilon$ describes the configuration of the evolution space as it gives positions of all sequences in an evolution instance. Whereby, momentum $p$ allows us to track time development of the evolution phase space. *FTEF* captures a snapshot of evolution by ascribing an imaginary mass to evolution engine $\epsilon$ and setting genetic velocity analogous to wobble $\omega_m$, whereby $\epsilon$ describes evolution effects on sequence homology and $\omega_m$ captures codon mutation as well as remnants of the evolution of the genetic code. Thusly, evolution momentum $p$ reflects change in sequence homology during gene evolution as well as the rate of mutation. By applying Newton's second law, we can also describe genetic acceleration of a DNA crossover thru the potential field as a derivative of mutation rate (12). Thusly, *FTEF* models phase space of an evolution instance and allows prediction of the trajectory of gene evolution by mapping the configuration space as a single point and tracking its trajectory across cycles.

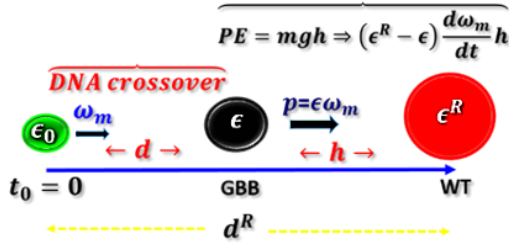 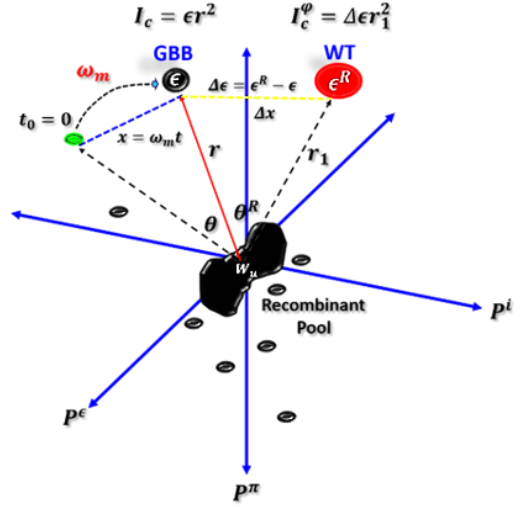

Fig. 2 Linear (A) and Rotation Models (B)

$$p = mv \Rightarrow \epsilon \cdot \omega_m \quad (11)$$
$$F = \sum ma \Rightarrow \sum \epsilon \cdot \frac{d\omega_m}{dt} \quad (12)$$

Work performed by the evolution force at DNA crossover junctions can be described by (13).

$$W = F \cdot d \Rightarrow \sum F \cdot \int (\omega_m + \omega_m^0) \, dt \quad (13)$$

To elucidate evolution dynamics, *FTEF* must describe the relative position of the parent sequence to the rigid body of DNA crossovers formed during evolution of the gene. As the initial position of the parental sequence within the evolution potential field cannot be ascertained, we solve for its relative position to the rigid body by viewing it as an ideal DNA crossover characterized by position vectors $\epsilon = 1$ and $\omega_m = 1$. We describe its relative position $(\epsilon^R, \omega_m^R)$ within the phase space by applying evolution weights $W_d$ and $W_p$ (4) and (5). Where, momentum $p$ of the WT sequence block thru evolution phase space is a function of position vector $\epsilon^R$ and mutation rate $\omega_m^R$ according to (14).

$$p = mv \Rightarrow \epsilon^R \cdot \omega_m^R \quad (14)$$

*FTEF* solves evolution potential energy $(V)$ as a steady state, where evolution potential is a function of the genetic distance of a DNA recombination from the position of the WT in configuration space (15). This genetic distance is described by potential mass $m_\varphi$ and distance $h$. Potential mass $m_\varphi = \epsilon^R - \epsilon$ is an imaginary mass characterizing the differential sequence homology $\Delta\epsilon$ remaining between the GBB and parental sequence after DNA recombination. $m_\varphi$ is dynamic as it is solved by comparing position vectors $(\epsilon, \epsilon^R)$ in phase space, thusly changes with sequence homology during the evolution process. Displacement $h = x^R - x$ describes distance of the DNA crossover instance to the WT sequence within the evolution potential field. Where, $x^R$ is the relative genetic step of the WT sequence block at $t_0$ and describes the relationship between its protein and DNA position vectors, thusly $x^R = 1$. Position vector $x$ describes the time-independent genetic step as described in (6). Evolution potential energy increases with genetic distance due to larger potential mass $m_\varphi$ and distance $h$ between particles **Fig. 2** (A). Thusly, configuration space $V$ is dynamic and changes as a function of differential sequence homology and evolution rate with idyllic DNA crossovers having smaller energies.

$$V = \sum m_\varphi \frac{d\omega_m}{dt} h = \sum a_\epsilon (x^2 - y) \quad (15)$$

$$\text{Where } a_\epsilon = \frac{m_\varphi}{t^2}, y = xx^R$$

We also solved potential energy as a function of evolutional acceleration $a_\epsilon$ through the potential field (15) right hand side of equality symbol. Where, position vector $y$ is the product of vector $x$ and $x^R$ and was derived from a polynomial derivation of their momentums in phase space.

The potential energy vector also allows comparison of gene sequence spaces in respect to their mutation rates. Where, the relationship between wobble and incremental potential energy changes within a recombinant pool may be described by a first order differential equation $dV = 2a_\epsilon x dx$. Where, vector $\vec{V}$ characterizing the gene's evolution is described in (16).

$$\vec{V} = \sum 2a_\epsilon x dx \quad (16)$$

Kinetic energy ($T$) configuration space captures magnitude of evolution force applied on a sequence space as force $F = \nabla T$ is the gradient of the energy scalar field. GBBs have high kinetic energy, thusly are defined as sequences displaying high momentums through phase space. They are characterized by a high degree of sequence conservation accompanied by a large magnitude of wobble indicating the introduction of genetic diversity at the DNA level.

$$T = \frac{1}{2}\sum \epsilon \cdot \omega_m^2 \Rightarrow \frac{1}{2}\sum a_\epsilon x \quad (17)$$

The Hamiltonian reflects evolutionary advantageousness of the sequence space (18). H configuration space describes evolution force applied on DNA crossovers over evolutional history of the sequence block as well as genetic distances of DNA crossovers to WT.

$$H = T + \overbrace{\sum mgh}^{V} \Rightarrow \frac{1}{t^2}\sum \left\{\frac{1}{2}\epsilon x^2 + m_\varphi(x^2 - y)\right\} \quad (18)$$

Time-independent incremental changes in the systems total energy may be described by a first order differential equation (19). Time-dependent wobble effects are described in (20). Vector $\vec{E}(t)'s$ configuration space describes incremental changes of total energy that result from changes in mutation rate respective to the phylogenetic history of the gene.

$$\vec{E} = 2\sum a_\omega (\epsilon + 2m_\varphi) dx \quad (19)$$
$$\vec{E}(t) = -4\sum J_\omega (\epsilon + 2m_\varphi) dx dt \quad (20)$$

The Lagrangian $\mathcal{L}$ of the system describes the path of the least evolutional resistance. The Lagrangian also allowed us to derive the motion equation as a function of position vectors $x, y$ (21). The optimal path for gene formation is enumerated by summation of the Lagrangians characterizing DNA crossovers occurring within each genomic alphabet forming the gene's DNA secondary code (DESC).

$$\mathcal{L} = \sum T - V \Rightarrow a_\epsilon \sum \left\{\frac{1}{2}Ax^2 + y\right\}, where \ A = \frac{\epsilon}{m_\varphi} \quad (21)$$

The motion equation of the phase space was also derived from its Lagrangian as given in (22), where $\dot{x} \equiv \omega_m$.

$$\frac{d}{dt}\frac{\partial \mathcal{L}}{\partial \dot{x}} - \frac{\partial \mathcal{L}}{\partial x} \Rightarrow a_\epsilon A\dot{x} - a_\epsilon Ax \quad (22)$$

State $S$ describes the evolutional equilibrium of the phase space and is solved as $\int \mathcal{L} dt$, thusly defines sequence space under the evolution curve with less negative states indicating highly evolvable spaces.

$$S = \int_{t_0}^{t_f} \mathcal{L} dt \Rightarrow \int_{t_0}^{t_f} \left(a_\epsilon \sum \left\{\frac{1}{2}Ax^2 + y\right\}\right) dt = -v_\epsilon \sum \left\{\frac{1}{2}Ax^2 + y\right\}, where \ A = \frac{\epsilon}{m_\varphi} \ and \ v_\epsilon = \frac{m_\varphi}{t} \quad (23)$$

### D. Analyzing Evolution Force Utilizing the Rotation Model

The Rotation Model (RM) analyzes evolution force $\tau_\varepsilon$ as a function of GBB moments of inertia about a rigid body of DNA crossovers that characterizes phylogenetic history of the sequence. Positions of particles in the configuration space are given by evolution engine $\varepsilon$ and their relative position to the rigid body described by standard deviation $r$. Each identifier $\varepsilon \in \{\epsilon, \omega_m, P^i, P^\pi\}$ acts as an engine that produces a driving force $\tau_\varepsilon$ and has its own configuration space. The RM allows us to estimate contributions of identifiers to genomic building block formation and due to linearity of FTEF functions also allows for multidimensional analysis.

The Langevin of the system is a function of the gradient of the Hamiltonian (24), where driving force $\tau_\varepsilon$ is the torque about the rigid body and evolutional decay $\zeta_\varepsilon = \nabla_x \times H$ is the curl of the scalar field and describes dissipation of information out of the system as evolutional noise. $\tau_\varepsilon$ is a function of evolutional moments of inertia $I_\varepsilon = mr^2 \to \varepsilon r^2$ of DNA crossovers and their acceleration $d\omega_m/dt$ thru the evolution potential field. Evolutional pressure $P_\varepsilon = \nabla_x H/sspace_r$ describes force applied on the sequence, where $sspace_r$ is the size of the ortholog-paralog sequence space.

$$\nabla_x H \to \tau_\varepsilon + \zeta_\varepsilon \quad (24)$$
$$\tau_\varepsilon = \sum I_\varepsilon \cdot \frac{d\omega_m}{dt} \quad (25)$$

Work performed by the evolution force is given by (26), where $\theta = \frac{1}{r}\int(\omega_m + \omega_m^0) dt$ describes the genetic step of the DNA crossover toward the parent sequence.

$$W = \sum \tau_\varepsilon \cdot \theta \quad (26)$$

The system's Hamiltonian is a function of rotational kinetic $T$ and potential energies $V$ about the rigid body. Rotational potential energy is a function of inertial vector $I^\varphi = m_\varphi r^2$ characterizing potential moments about engine $\varepsilon$ and describes evolution potential in respect to the phylogenetic relationship between recombinant and WT sequence positions in evolution phase space, captured by potential mass $m_\varphi = \epsilon^R - \epsilon$. It gives standard deviation of the DNA crossover from the rigid body, thusly describes convergence and divergence of sequences to the WT in respect to the phylogenetic history of the sequence.

$$H \to \overbrace{\frac{1}{2}\sum I_\varepsilon \cdot \omega_m^2}^{T} + \overbrace{\frac{1}{2}\sum I^\varphi \cdot \omega_m^2}^{V}, \quad (27)$$

Configuration state $S$ is a function of the difference in DNA crossover angular momentum $L_T$ in direction of the kinetic energy vector and its angular momentum $L_V$ in direction of the potential energy vector. Reactions characterized by equilibria in the kinetic energy direction are evolutionarily favorable. Configuration state $S$ is also a function of position vectors $(x, x^R)$ that describe convergence or divergence of the DNA crossover to the WT sequence block and the relative distance of the parent sequence to the rigid body. The function $f(x, x^R)$ reflects the hierarchical relationship between protein and DNA position vectors in the phase space.

$$S = \int_{t_0}^{t_f} \mathcal{L} dt \Rightarrow \int_{t_0}^{t_f} (T - V) dt \qquad (28)$$
$$= -\frac{1}{2} \sum [L_T - L_V] \cdot f(x, x^R)$$

Incremental changes in total energy $d\vec{E}$ of the configuration space in respect to mutation rate are described in (29), and incremental changes to its equilibria $d\vec{S}$ described in (30).

$$d\vec{E} = \sum \tau_\varepsilon dx, \qquad (29)$$
$$d\vec{S} = -\frac{1}{2} \sum L_\varepsilon \, dx, \qquad (30)$$

*E. Evaluating Synthetic Structures*

The *FTEF* defines wobble as the conservation of structure in face of genetic diversity. When wobble occurs at macroscopic levels and higher, it is referred to as structural. An example is phyllotaxis, the arrangement of leaves and the deformation configurations seen on plant surfaces described in [42]. These Fibonacci-like patterns are conserved across plant species that encompass a broad range of genetic diversity, thusly according to *FTEF* display structural wobble. *FTEF* solves for wobble as a conditional probability of target structure similarity to native states. The probability that state $x_s$ formed during synthetic evolution will share homology with the native state is a function of closeness probability $\theta_i$, where $i \in P$ comprises physiochemical properties volume, hydrophobicity, charge and folding propensity.

$$wobble = f(x_s | \theta_i), \ \ where, i \in P \qquad (31)$$

To prevent structural perturbations, SYN-AI performs high-resolution pattern recognition by analyzing discrete sequence spaces occurring across protein structures and walking GBB protein sequences in single steps of one residue. Each step comprises a discrete group of three residues and overlaps the previous step. Where propensity of characteristic (*i*) within the sequence space is summated as illustrated in (32). Structural propensity ($p$) occurring within a discrete sequence space is fingerprinted by probability density function ($\delta$) as illustrated in (33). Area under the density curve $\int p \, dp$ is normalized by partition function $\sigma$ describing summation of characteristic $i$ across the structure. This allows SYN-AI to characterize the taste of discrete sequence spaces. Proteins are characterized by diverse flavors describing small changes in physiochemical properties occurring both locally and globally. Closeness of the synthetic structure to the native is described by probability $\theta_i$ and solved as a function of synthetic $\delta_i^{syn}$ and native $\delta_i^{nat}$ states described in (34).

$$p = \sum_n^{sppace} \sum_i AA_n^i, \ \ where \ i \in P \qquad (32)$$
$$\delta = \frac{1}{\sigma} \int p \, dp, \qquad (33)$$
$$\theta_i = 1 - \frac{|\delta_i^{syn} - \delta_i^{nat}|}{\delta_i^{nat}} \qquad (34)$$

In solving probability of structural state $x_s$, $\theta_i$ is factored across $n$ sequence spaces comprising the structure. Where, $i$ is an element of S: {secondary, super secondary and quaternary} structural groups.

$$\prod_{i=1}^{sspace} x_s = \theta_1 \times \theta_2 \cdots \times \theta_{sspace}, where \ i \in S \qquad (35)$$

*FTEF* solves for structural wobble as a function of average closeness $\langle Closeness \rangle$ of synthetic and native states. Where, probability $\theta_i$ is summated over $n$ discrete sequence spaces comprising the structure and characteristic (*i*). N reflects the total number of measurements and $i$ is an element of set $P$. Alternatively, $\langle Closeness \rangle$ may be more precisely described by probability $\prod x_s$. Structural wobble is solved as a function of $\langle Closeness \rangle$ and protein similarity $Prot_s$.

$$\langle Closeness \rangle = \frac{1}{N} \sum_i \sum_n [\theta_i]_n, where \ i \in P \qquad (36)$$
$$wobble = \frac{\langle Closeness \rangle}{Prot_s} - 1 \qquad (37)$$

III. EXPERIMENTAL METHODS AND PROCEDURES

*A. High Performance Computing*

SYN-AI experiments were performed on the Stampede 2 supercomputer located at the Texas Advanced Computing Center, University of Texas, Austin, Texas. Experiments were performed in the normal mode utilizing SKX compute nodes comprising 48 cores on two sockets with a processor base frequency of 2.10 GHz and a max turbo frequency of 3.70 GHz. Each SKX node comprises 192 GB RAM at 2.67 GHz with 32 KB L1 data cache per core, 1 MB L2 per core and 33 MB L3 per socket. Each socket can cache up to 57 MB with local storage of 144 /tmp partition on a 200 GB SSD.

*B. Simulating DNA Crossovers*

SYN-AI simulated evolution by partitioning the parental *Bos taurus* 14-3-3 ζ gene into a DNA secondary code DSEC comprised of 34 genomic alphabets and performing $1 \times 10^9$ DNA crossovers within genomic alphabets comprising the DSEC. DNA hybridizations were performed at 19, 20 and 21 base pairs allowing us to capture mutations in three open reading frames. DNA hybridization partners were randomly selected across an orthologue/paralogue sequence space constructed by an automated NCBI-Blast. The sequence space comprised of $2.5 \times 10^6$ bp of genetic material and genes at a homology threshold of > 80 percent identity to parental *Bos taurus* 14-3-3 ζ. DNA hybridizations were simulated in 3 mM

Mg$^{2+}$ and 1.2 mM dNTP at 328.15° kelvin [14]. Gibb's free energy was calculated according to [14] and a penalty assessed for DNA base pair mismatches.

### C. Simulating Natural Selection

Selection was limited to thermodynamically favored DNA crossovers utilizing an inverse tangent sigmoidal function to scale Gibb's free energy vectors. Free energy vectors were converted to Heaviside nodes by applying an experimental bias and by subsequent transformation utilizing a sinc (x) function in conjunction with a Boolean function. Sequences generating a signal of 1 were considered as GBB candidates and passed thru a cascade of subsequent neural networks. A second round of natural selection utilized pattern recognition filters to remove sequences characterized by long stretches of low sequence homology, thusly lowering the probability of protein perturbations. A third round of selection limited DNA crossovers to those comprised of evolutionarily favored mutations based upon Blosum 80 mutation frequency. In a fourth round of selection, DNA crossovers were limited to those displaying evolution earmarks characterized by (+) molecular wobble vectors [43]. In a final round of natural selection, DNA crossovers were limited to those characterized by a high magnitude of evolution force.

### D. Engineering Synthetic Super Secondary Structures

Super secondary structures were identified using STRIDE knowledge based secondary structure algorithms [44], which converted the 14-3-3 ζ docking protein sequence to a DNA tertiary code (DTER). Synthetic motifs were engineered by ligating GBBs randomly selected from genomic alphabet libraries encompassing 5' to 3' terminals of parental structures. Synthetic super secondary structures were equilibrated with native structures using a cleaving algorithm to remove 5' and 3' prime overhangs. Natural selection was limited to synthetic structures characterized by naturally occurring mutation based on BLOSUM80 mutation frequency. A final round of natural selection imposed a secondary structure homology threshold of > 88 percent identity to parental 14-3-3 ζ using a PSIPRED 4.0 [45], [46] based algorithm to evaluate secondary structure. Synthetic structures that passed natural selection were stored in DTER libraries for writing DNA code.

### E. Engineering 14-3-3 ζ Docking Genes

14-3-3 ζ docking genes were engineered by walking the DTER followed by random selection and ligation of synthetic super secondary structures stored in DTER libraries. SYN-AI constructed a library of $1 \times 10^7$ simulated genes that were passed thru a set of neural networks that evaluated closeness of synthetic structural states to native states as described in the '**Theory**' section, with a minimal closeness threshold of > 90 percent identity. Selection was limited to proteins comprised of a high composition of naturally occurring mutations based on their average BLOSUM80 mutation frequency. A further round of natural selection restricted selection to synthetic 14-3-3 ζ docking proteins having secondary structure identities located within the top quantile of normalized vectors [43]. A final round of selection enriched for functional 14-3-3 ζ docking proteins by comparing synthetic protein active sites and hydrophobic interfaces to those of native *Bos taurus* 14-3-3 ζ, a closeness threshold of > 90 percent identity was set.

## IV. RESULTS AND DISCUSSION

### A. Analysis of Evolution Force

We validated *FTEF* by proof of concept, whereby synthetic evolution artificial intelligence SYN-AI engineered a set of three 14-3-3 ζ docking proteins. To simulate evolution, the parental *Bos taurus* 14-3-3 ζ docking gene was partitioned into DNA secondary (DESC) and tertiary (DTER) codes based on hierarchical protein structures. We identified genomic building blocks by performing DNA hybridizations within the DSEC and by applying natural selection. Whereby, evolution force associated with DNA crossovers was calculated using Linear and Rotation models as described in the 'Theory' section. To fully enumerate genomic building blocks back to LUCA, we calculated evolution force across single and multidimensional configuration spaces as described in the 'Supplemental Materials'. Following identification of GBBs, the DNA tertiary code was used as a template for engineering synthetic super secondary structures. Synthetic structures that passed natural selection based on BLOSUM80 mutation frequency and the closeness of native and synthetic secondary structures were placed in DTER libraries. SYN-AI engineered a set of 10 million docking proteins by walking the DTER and randomly ligating synthetic structures stored in DTER libraries. Notably, this large gene set was reduced to three structurally conserved 14-3-3 ζ docking proteins by simulating natural selection.

Linear Model configuration spaces were characterized by broad distributions of evolution force and low resolution of GBBs, however they captured formation of multiple evolution foci suggesting successful simulation of the time-development of the evolution phase space. Sequence 1 was characterized by DNA crossovers distributed around the population expectation at ($\omega_m = 0, \epsilon = 1.0$) Fig 3A (i). However, sequence 2 was characterized by localization of GBB foci in positive and negative evolution phase space indicating presence of strong selection and deselection pressures and a change of biological function Fig. 3A (ii). Convergence toward WT was signaled by localization of a hotspot at ($\omega_m = 0.45, \epsilon = 1.6,$). Notably, localization of a GBB hotspot in (-) phase space at ($\omega_m = -0.5, \epsilon = 0.3$) reflects deselection and subsequent speciation that resulted in the change of biological function. Relaxation of evolutional stringency was corroborated by the decrease in evolutional conservation from the population expectation of $\epsilon = 1.0$ to $\epsilon = 0.3$, Fig.3A (ii). The pattern of light blue GBB distributions leading to the foci show the time evolution of the sequence block and mutations that lead to the function change. Nonrandom concentric distributions of GBB indicate that the *FTEF* simulated evolution of the sequence, while concentric yellow hotspots located around foci indicate parallel evolution that resulted in structures of similar function as confirmed by sequence alignments performed in [43]. The two less prominent foci localized at ($\omega_m = 0, \epsilon = 0.6$) and ($\omega_m = -0.1, \epsilon = 1.4$) indicate the involvement of additional evolution mechanisms. Contrarily, the near normal distribution of DNA crossovers in sequence 1 is due to genetic dispersion resulting from neutral evolution, whereby evolutional noise prevented foci formation.

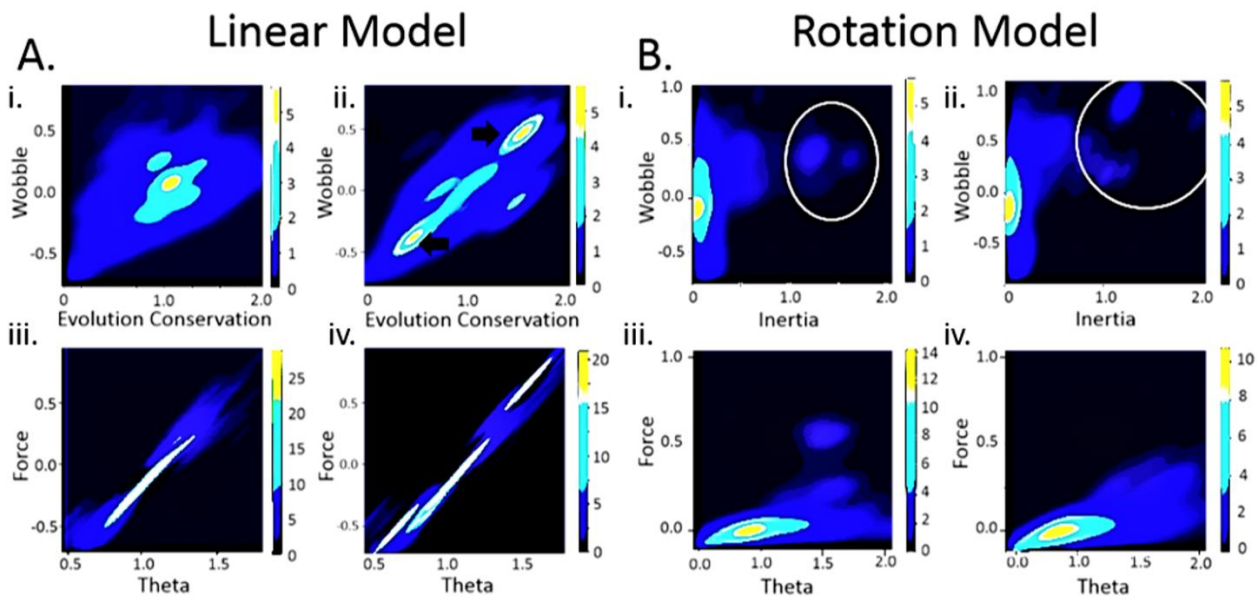

Fig. 3 Evolution Force Linear vs. Rotation Model.
Linear Model (A). Evolution force distribution sequence space 1 of the DSEC code (i). Evolution force distribution sequence space 2 (ii). Work distribution sequence space 1 (iii). Work distribution sequence space 2 (iv). Rotation Model (B). Evolution force distribution sequence space 1 (i). Evolution force distribution sequence space 2 (ii). Work distribution sequence space 1 (iii). Work distribution sequence space 2 (iv).

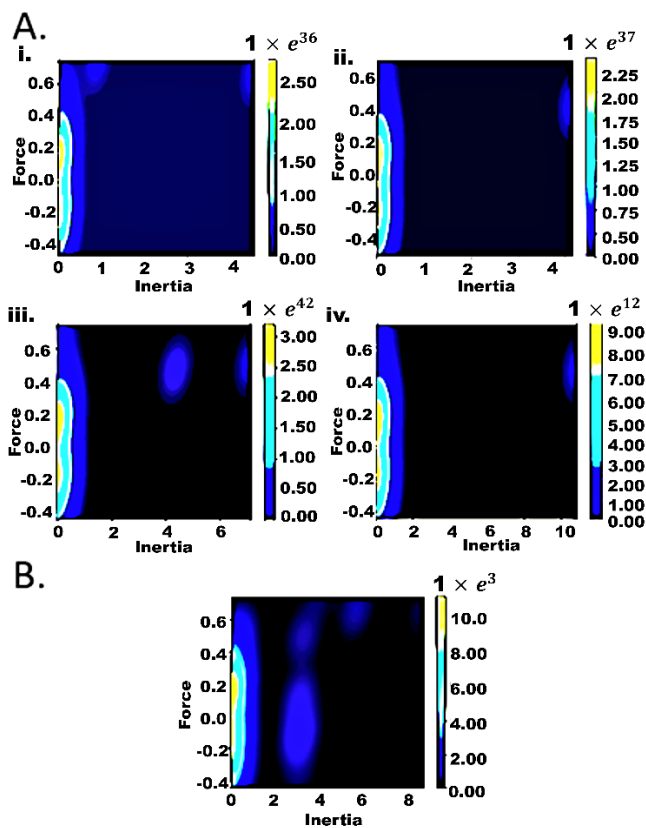

Fig. 4 Evolution Force Distribution across Multidimensional Planes

Evolution configurations simulated by the Rotation Model achieved high-resolution of GBBs denoted by circles, Fig. 3B. While SYN-AI employed both methods in GBB identification, the RM was predominantly used in neural networks. The RM did not capture formation of multiple foci in sequence spaces 1 and 2. However, it captured formation of dual foci in three-dimensional phase spaces alpha Fig. 4A (i), beta Fig. 4A (ii), gamma Fig. 4A (iii), and rho Fig. 4A (iv) as well as four-dimensional phase space Fig. 4B, which show formation of hotspots in positive and negative phase space and captures divergence of function during their time-development. Linear and Rotation models detected evolution mechanisms at different sensitivities, thusly their combined use allowed for detailed investigation of the evolution of sequence blocks. When using the RM, formation of GBBs in (+) evolution phase space was characterized by foci comprised of sequences that display a high magnitude of evolution force and inertia. Results were consistent across configuration spaces, whereby increased dimensionality improved GBB resolution.

Notably, gene sequence spaces exhibited different behaviors due to thermodynamic barriers that form during evolution as a result of speciation. Sequences retaining high homology to the ancestor bind more stably in DNA hybridizations due to higher magnitude Gibb's free energies and lower thermodynamic penalties. Thusly, none of the GBB phase spaces were similar with all displaying unique time developments due to different rates of speciation. If these processes were random and due to random hybridization of DNA fragments within gene sequence space, the phase spaces would display a similar distribution of particles. Thusly, we corroborate that *FTEF* captures boundary conditions governing gene evolution and depicts independent pathways of phase space evolution. Whereby, we hypothesize that Gibb's free energy partitions that separate sequence phase spaces may have guided evolution processes and are intrinsic components of the evolution force and gene evolution.

We corroborated that the evolution force is a low energy system as work performed in positive and negative directions eliminated each other, Fig. 3 (iii, iv). Work $W = \sum \nabla_{x,t} E \cdot \theta$ performed by the evolution force is a function of the energy gradient $F = \nabla_{x,t} E$ and genetic displacement $\theta$, thusly it is a function of the slope of the energy field and dependent on the evolvability of the sequence space. Its dependence on $\nabla_{x,t} E$ means it is a superimposition of the Hamiltonian, thusly is not static but dynamic as kinetic and potential energy landscapes are in constant flux. Work configuration space evolves during DNA recombinations and time-development is characterized by changes in genetic distance $x$. Work associated with GBB formation was distributed around a genetic distance of $\theta = 1.0$ which is the population expectation. While there was skewing of its mean distribution to (-) phase space, it was counterbalanced by sparse occurrences of sequences in (+) phase space. We expected work to be significantly skewed toward (-) evolution space due to random hybridization of non-homologous DNA sequences. Offset of work in positive and negative phase space suggests that selection protocols implemented by the *FTEF* are very reliable. Notably, our experiments corroborate findings of Aravind [47] and capture a complex interplay of evolution conservation and genetic diversity during gene evolution.

*B. Analysis of Synthetic 14-3-3 ζ Docking Proteins*

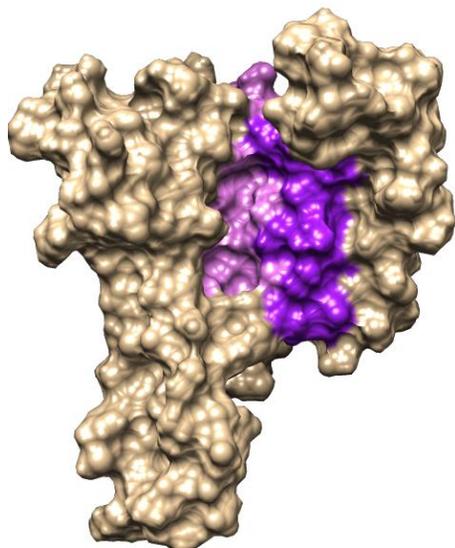

Fig. 5 SYN-AI-1 ζ Structure. SYN-AI-1 ζ three-dimensional structure was estimated using I-Tasser (Zhang Laboratory, University of Michigan). Residues 99 – 129 are colored purple and residues 152 – 180 are colored cyan.

The *FTEF* captured effects of fitness and evolution rate on discrete sequence blocks as well as provided information on protein domain formation. While we performed $1 \times 10^9$ DNA crossovers within the 14-3-3 ζ DNA secondary code and simulated $1 \times 10^7$ proteins none of the three docking proteins that passed natural selection contained mutations between residues 99 – 129 and 152 – 180 with exception of a $I \rightarrow S$ at residue 106 of SYN-AI-1 ζ and SYN-AI-3 ζ and a $Y \rightarrow S$ at residue 179 of SYN-AI-1 ζ and SYN-AI-2 ζ as we reported in [43]. These regions were almost fully conserved suggesting that they are critical to fitness and are characterized by very slow evolution rates. When we superimposed these sequences to the SYN-AI-1 ζ three-dimensional structure they localized to the amphipathic groove, Fig. 5.

The amphipathic groove has been reported to be critical to protein function and is the location of the 14-3-3 ζ active site as well as BS01, BS02 and BS03 ligand binding sites [43], [48]. The ability of *FTEF* to simulate natural selection was corroborated by the positioning of conserved sequence blocks in synthetic 14-3-3 ζ three-dimensional structures. The highly conserved sequence blocks are separated by 23 residues on the protein primary sequence, however, when mapped to the 14-3-3 ζ structure they are located adjacent to each other within the amphipathic groove with overlapping Van der Waals surfaces. The spatial configuration of these sequence blocks suggest that they evolved as separate domains and that in addition to their contribution to the active site and ligand binding, they may also play additional functional roles. When, we invert the structure we notice that residues 130 – 151 located between the conserved sequence blocks are associated with the spine of the protein, Fig. 6 (A). The spine allows flexibility when performing bend and flex mechanisms during communication between the 14-3-3 ζ active site and C' terminal helix H3 tail. Although this role is critical to function our data suggest this region can tolerate mutation. The highly conserved sequences play a dual role in protein activity and flexibility allowing the protein to capture the ligand and change configuration to the 'closed' conformational state. We form this hypothesis based upon the position of these residues Fig. 6 (A). Additionally, ribbon structures depicted in Fig. 6 (B) corroborate that the sequences evolved as separate motifs.

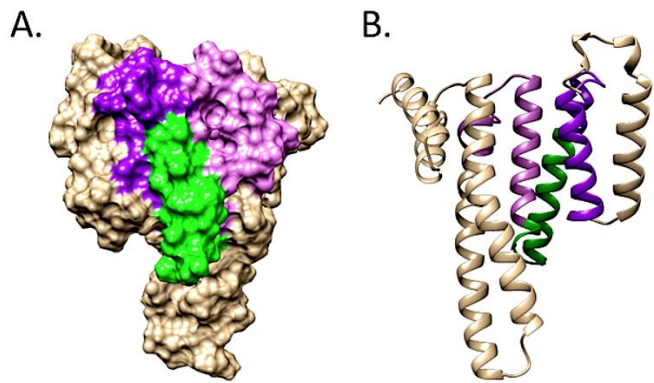

Fig. 6 SYN-AI-1 ζ Structure Reverse View. Residues 99 – 129 (purple), residues 152 – 180 (cyan), and residues 130 – 151 are colored (green). Surface structure (A). Ribbon structure (B).

According to Ghosh, cooperative communications between protein domains is a critical component of protein function

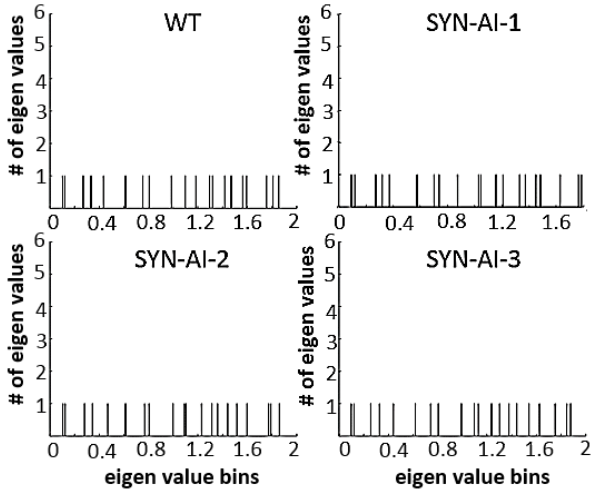

Fig. 7 Normal Mode Analysis. Eigenvalues of the native *Bos taurus* 14-3-3 ζ monomer and synthetic proteins SYN-AI-1 ζ, SYN-AI-2 ζ, and SYN-AI-3 ζ.

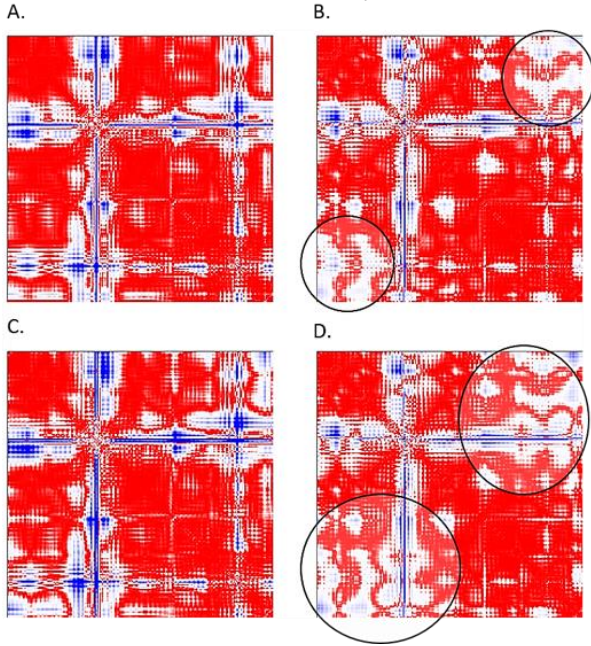

Fig. 8 Distance Matrices. Normal mode 7 vibrational dynamics of native *Bos taurus* 14-3-3 ζ and synthetic docking proteins were evaluated utilizing the anisotropic network model. Native *Bos taurus* 14-3-3 ζ (A), synthetic docking proteins SYN-AI-1 ζ (B), SYN-AI-2 ζ (C), and SYN-AI-3 ζ (D).

[49]. Cooperative communications within synthetic docking proteins were analyzed using the anisotropic network model, ANM2.1 [50]. Based on predicted eigenvalues Fig. 7, despite a significant sequence divergence of 7.33 achieved by *FTEF* the global allosteric footprint was conserved. However, altered locations of modes suggests altered low frequency vibrations as well as rewiring of cooperate communications within the docking protein, Fig. 7. For instance, near the eigenvalue of 0.2 in the native protein there are two associated modes, however in SYN-AI-1 ζ there is a 3rd closely associated vibrational mode as the mode near the eigenvalue 0.4 shifted leftward. Likewise, near the 1.0 eigenvalue in the native 14-3-3 ζ docking protein we observed one vibrational mode, however, in SYN-AI-1 ζ and SYN-AI-2 ζ a second mode was introduced to the motion. Furthermore, there was an obvious change in motion involving the three modes located near the eigenvalue of 1.8 in all three synthetic docking proteins.

We further analyzed for changes in cooperative motions within docking protein monomers by comparing intra-residue distance fluctuations occurring during normal mode 7, Fig. 8. Native and synthetic distance matrices overlapped well, thusly corroborating that synthetic evolution by *FTEF* achieved global conservation of 14-3-3 ζ architecture and vibrational dynamics. The ability of *FTEF* to engineer proteins without disrupting normal modes is critical as 14-3-3 ζ participates in over 230 protein-protein interactions and numerous signal transduction pathways. Notably, while the global vibrational footprint was conserved, local distance variations denoted by circled areas suggests that *FTEF* achieved pathway specific rewiring of cooperative communications.

V. CONCLUSION

In the current study, we validated the 'Fundamental Theory of the Evolution Force: *FTEF*' by proof of concept. Whereby, a synthetic evolution artificial intelligence (SYN-AI) was used to engineer a set of three architecturally conserved 14-3-3 ζ docking genes with the *Bos taurus* 14-3-3 ζ gene serving as a template for time-based DNA codes to guide the engineering process. Notably, *FTEF* allowed us to observe evolution force associated with genomic building block formation as well as to observe speciation processes. The theory also allowed us to observe gene convergence and divergence over an evolution phase space going back to LUCA. Importantly, we were able to introduce significant genetic diversity into docking proteins while conserving global and local protein architecture as well as vibrational modes. This is significant as 14-3-3 ζ docking proteins play significant roles in cancer and neurodegenerative disease, whereby synthetic evolution by *FTEF* may offer an opportunity for novel drug discovery by possibly modulating ligand interactions and signal transduction pathways.

APPENDIX

*A. Supplemental Information*

1. Force and Energy Dynamics in Two-dimension Planes of Evolution

Multidimensional analysis of evolution force associated with genomic building block formation was performed using the "Rotation Model" as a function of moments of inertia about selectivity states $p_\epsilon$, $p_\omega$, $p_i$, and $p_\pi$. Selectivity states characterize evolutional fitness of a DNA crossover in respect to the evolution engine and are calculated by distributing GBB moments of inertia $I^\varepsilon$ over the summation of inertial moments comprising the rigid body. The inertial moment $I_{p^\varepsilon}$ about evolution engine $\varepsilon$ is then solved by setting the selectivity state analogous to mass and multiplying by variance $\sigma_{p^\varepsilon}^2$ from the rigid body. Where, $\varepsilon$ is an element of the four fundamental evolution engines (evolutionary conservation, wobble, DNA binding state, periodicity) and the rigid body characterizes full enumeration of DNA crossovers occurring in sequence space ($sspace^r$). Thusly, $I_{p^\varepsilon}$ characterizes moments of inertia about the evolution engine respective to its phylogenetic history back to LUCA. Evolution force $\tau_\varepsilon = \sum I_{p^\varepsilon} \cdot a$ is solved as a function of inertia about the evolution engine and the angular acceleration which is the derivative of the mutation rate $\omega_m$.

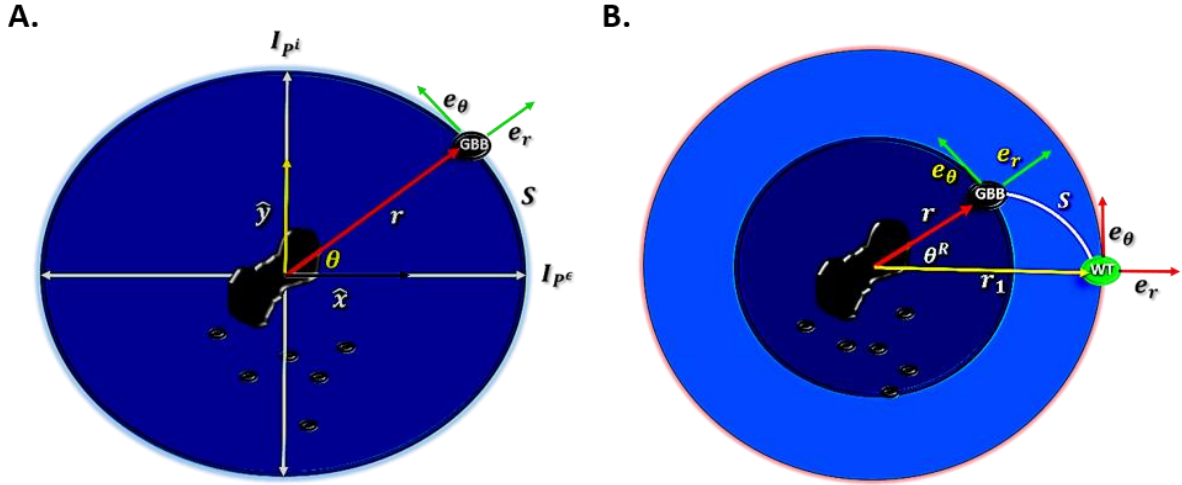

Fig. 9 Identification of GBB Formation and Force Distribution in Two-dimensional Evolution Planes

$$I_{p^\varepsilon} = m \cdot r^2 \Rightarrow \overbrace{\left[I^\varepsilon / \sum_{n=1}^{sspace^r} I^\varepsilon\right]}^{selectivity} \cdot \sigma_{p^\varepsilon}^2 \quad (38)$$

$$Where, \varepsilon \in \{four\ fundamental\ evolution\ engines\ \}$$

While a simple solution of two-dimensional evolution force is given by $\tau_\varepsilon = \sum I_{p^\varepsilon} \cdot a$, we can solve inertia $I$ in $\mathbb{R}^2$ as the resultant of inertial vectors in orthogonal directions with each characterizing an evolution engine, thusly there are six non-redundant inertial vectors formed in $\mathbb{R}^2$ evolution space.

$$I_\alpha = I_{p^i} + I_{p^\epsilon} \quad (39)$$
$$I_\beta = I_{p^i} + I_{p^\omega} \quad (40)$$
$$I_\gamma = I_{p^\omega} + I_{p^\epsilon} \quad (41)$$
$$I_\rho = I_{p^\pi} + I_{p^\omega} \quad (42)$$
$$I_\sigma = I_{p^\pi} + I_{p^\epsilon} \quad (43)$$
$$I_\tau = I_{p^\pi} + I_{p^i} \quad (44)$$

The Rotation Model describes evolution force occurring within $\mathbb{R}^2$ as depicted in Fig. 9. Where, GBB instances are characterized as DNA crossovers resulting in high moments of inertia about the rigid body in $\mathbb{R}^2$. Displacement $r$ from the rigid body is the resultant of moments of inertia in $(x, y)$ directions and orthogonal evolution engines. Unit vectors $e_\theta$ and $e_r$ are functions of displacement vectors $(\theta, r)$ comprising the rigid body divided by their magnitude Fig. 9 (A). Relative position of the parental sequence in the evolution potential field is described by radius $r_1$ Fig. 9 (B). Whereby, potential energy of the system is a function of genetic distance S characterizing the genetic step from the GBB to the parental sequence and relative displacement $\theta_R$. The Hamiltonian and Lagrangian are a function of inertial kinetic and potential energies generated about the described orbitals Fig. 9 (B). Where, evolution force $\tau_\varepsilon$ exerted in formation of a GBB is a function of evolutional torque $\tau_r$ applied about the rigid body on fulcrum $r$ as well as angular momentum $L_S$ and torque $\tau_S$ of the DNA crossover about displacement vector $S$ as described in (45).

$$\tau_\epsilon = \sum I \cdot a \Rightarrow \sum (\tau_r + L_S \theta)\hat{e}_r + (\tau_S)\hat{e}_\theta \quad (45)$$

Kinetic energy $(T)$ of the phase space is characterized by a polynomial function describing its distribution about fulcrums $r$ and $S$. Unit vectors $e_\theta$ and $e_r$ describe the evolutional center as they are normalized expected positions of evolution vectors.

$$T = \tfrac{1}{2}\sum I \cdot \omega_m^2 \quad (46)$$
$$\Rightarrow \sum T_r(\hat{e}_r^2) + 2T_{rS}(\hat{e}_r\hat{e}_\theta) + T_S(\hat{e}_\theta^2)$$

We express the system's Lagrangian $\mathcal{L}$ as the difference in two polynomial functions that describe kinetic energy about fulcrums $r$ and $r_1$, Fig. 9 (B). Where, radius $r_1$ describes the relative distance of the parent sequence to the rigid body and radius $r$ is the distance from the DNA crossover to the rigid body. $\mathcal{L}$ is also a function of unit vectors $(\hat{e}_r, \hat{e}_\theta)$ that describe expected linear and rotational evolution distances in respect to the rigid body.

$$\mathcal{L} = T - V \quad (47)$$
$$\Rightarrow \tfrac{1}{2}\sum(I\omega_r^2 + 2I\omega_r\omega_S + I\omega_S^2) \cdot f(\hat{e}_r, \hat{e}_\theta)$$
$$- \tfrac{1}{2}\sum(I^\varphi \omega_{r_1}^2 + 2I^\varphi \omega_{r_1}\omega_S + I\omega_S^2) \cdot g(\hat{e}_r, \hat{e}_\theta)$$

The motion equation of the evolution configuration space is described by (48), where $\dot{x} \equiv \omega_m$ gives the mutation rate and $x$ is the genetic step.

$$\frac{d}{dt}\frac{\partial \mathcal{L}}{\partial \dot{x}} - \frac{\partial \mathcal{L}}{\partial x} = 0 \quad (48)$$

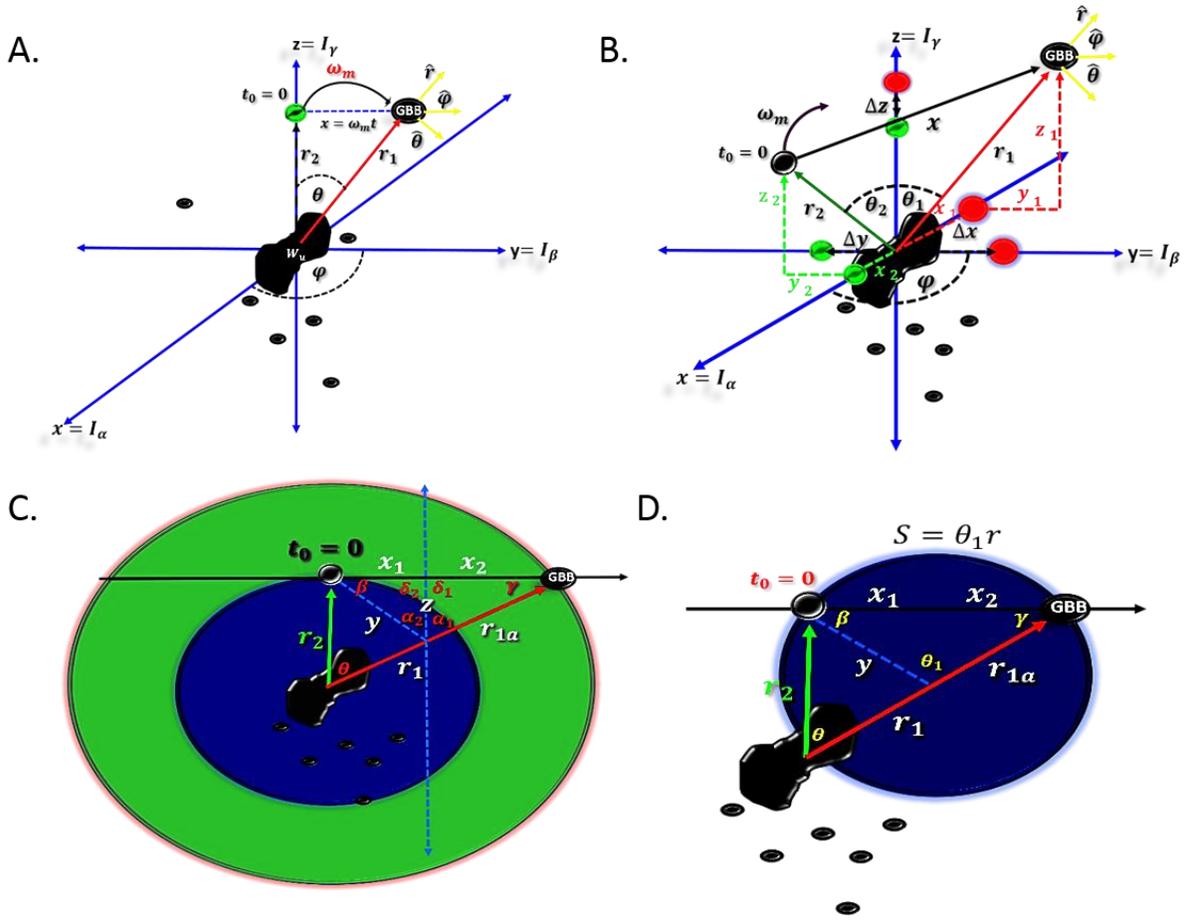

Fig. 10 Inertial Distribution in Three-dimensional Evolution Space

State $S$ of the configuration space is a function of its kinetic and potential state and depends on angular acceleration about the evolutionary axis.

$$S = \int_{t_0}^{t_f} \mathcal{L} dt \Rightarrow \qquad (49)$$

$$-\frac{1}{2}\Sigma\left[\overbrace{(r(L_r + 2L_S) + L_S S)\cdot f(\hat{e}_r, \hat{e}_\theta)}^{\text{kinetic state}} - \underbrace{\Sigma\overline{(r_1(L_r + 2L_S) + L_S S)\cdot f(\hat{e}_r, \hat{e}_\theta)}}_{\text{potential state}}\right]$$

2. Force and Energy Dynamics in Three-dimension Planes of Evolution

We modeled GBB formation in $\mathbb{R}^3$ as a DNA crossover at time $t = 0$ with genetic displacement toward the parental sequence. Where, genetic acceleration $d^2x/dt^2$ in $\mathbb{R}^3$ is a function of molecular wobble, angular displacement $\theta$, and the theoretical azimuth angle $\varphi$. Three-dimensional phase space can form as the resultant of two-dimensional inertial vectors described in the previous section. While there are several permutations we can perform, we show the formation of $\mathbb{R}^3$ evolution phase space as a function of inertial vectors $I_\alpha$, $I_\beta$, and $I_\gamma$, Fig. 10 (A). Time dependent displacement of position vector $x$ is illustrated in Fig. 10 (B), where genetic step $x = \Delta x + \Delta y + \Delta z$ gives the genetic distance between the expected and experimental position of the GBB in phase space at $t = 0$. Incremental inertial changes $\Delta x$, $\Delta y$ and $\Delta z$ are described by the distance between unit vectors $\hat{x}$, $\hat{y}$ and $\hat{z}$ (green) and actual positions (red) of GBBs in each inertial plane. The expected position of the GBB at $t = 0$ is the resultant of unit vectors $\hat{x}$, $\hat{y}$ and $\hat{z}$, while the experimental GBB position is the resultant of $x_1$, $y_1$, $z_1$ inertial positions in $I_\alpha$, $I_\beta$, and $I_\gamma$ inertial planes. $\hat{x}$, $\hat{y}$ and $\hat{z}$ characterize expected DNA crossover positions and are functions of the summation of inertial vectors occurring within the rigid body divided by their magnitude thusly reflect the population mean position. To solve for arc length $S$, we modeled the expected and experimental position of GBBs as particles orbiting the rigid body on different evolutionary paths, Fig. 10 (C). Arc length $S$ was solved by rotation about the inertial center and creating a midsection between distance vector $x$. This permitted re-centering of the inertial center and formation of right tringles that allowed elucidation of angles associated with distance vector $x$, arc length $S$, and angle $\theta_1$ utilizing the law of Sines, Fig. 10 (C). We reset the inertial center and restored the original relationships Fig. 10 (D).

Notably, by expanding force analysis to $\mathbb{R}^3$ configuration space, we identified unique genomic building blocks and not only expanded the GBB candidate pool, but also increased the probability of engineering functional genes. When considering distribution of evolution force in three dimensions, thirty non-redundant permutations of inertial vectors form as functions of two-dimensional inertial moments $I_\alpha, I_\beta, I_\gamma, I_\rho, I_\sigma$ and $I_\tau$. They formed both as resultants and dot products of evolution configuration spaces.

Additionally, three-dimension inertia vectors $I_\alpha, I_\beta, I_\gamma$, and $I_\rho$ were formed as resultants of selectivity states $I_{p^i}, I_{P^\omega}, I_{P^\epsilon}$ and $I_{P^\pi}$.

$$I_\alpha = I_{p^\epsilon} + I_{p^\omega} + I_{p^i} \tag{50}$$
$$I_\beta = I_{p^i} + I_{p^\epsilon} + I_{P^\pi} \tag{51}$$
$$I_\gamma = I_{p^i} + I_{p^\omega} + I_{P^\pi} \tag{52}$$
$$I_\rho = I_{p^\epsilon} + I_{p^\omega} + I_{P^\pi} \tag{53}$$

Four non-redundant permutations of three-dimension planes also form as dot products of inertial vectors characterizing the selectivity states of the four evolution engines.

$$I_{\alpha 1} = I_{p^i} \cdot I_{p^\epsilon} \cdot I_{p^\omega} \tag{54}$$
$$I_{\beta 1} = I_{p^i} \cdot I_{p^\epsilon} \cdot I_{P^\pi} \tag{55}$$
$$I_{\gamma 1} = I_{p^i} \cdot I_{p^\omega} \cdot I_{P^\pi} \tag{56}$$
$$I_{\rho 1} = I_{p^\epsilon} \cdot I_{p^\omega} \cdot I_{P^\pi} \tag{57}$$

We obtained solutions for evolution system dynamics in $\mathbb{R}^3$ including the evolution force vector $\vec{F}$, gradient $\vec{\nabla}\vec{F}$ as well as divergence $\vec{\nabla} \cdot \vec{F}$ and curl $\vec{\nabla} \times \vec{F}$ about the rigid body. This allowed analysis of evolutional proneness of genes and gene regions as well as for optimization of experimental conditions. The rotation model describes evolution force about a rigid body of particles characterizing the full enumeration of DNA recombinations over the evolutional history of the gene. The rigid body creates an evolutional gravitational field, whereby as described in [51] the force gradient $\vec{\nabla}\vec{F}$ gives a snapshot of collective directions of acceleration vectors and gravitational force fields. This allows us to analyze directional changes of evolution force within sequence phase spaces and to determine evolution engines that have greater impact on GBB formation under varying thermodynamic conditions. $\vec{\nabla}\vec{F}$ gives a snapshot of phylogenic dynamics of the configuration space, identifying gene regions that are more resistant or susceptible to mutation. Whereby, the dot product of the force gradient and mutation rate $\vec{\nabla}\vec{F} \cdot \omega_m$ gives the rate of change of the evolution force field during time development of the phase space. Divergence $\vec{\nabla} \cdot \vec{F}$ of the evolution force field gives a snapshot of evolution dynamics allowing comparison of configuration spaces by describing the separation of force field lines. We can capture the rate and direction of field expansion and contraction by the expression $\vec{\nabla} \cdot \vec{F} \cdot \omega_m$. Lastly, we evaluate curl of the force vector about the rigid body. This allows for the fine-tuning of experimental conditions by analysis of infinitesimal evolution force field rotations and evaluation of energy decay.

$$\text{Let } r = \|I_\alpha + I_\beta + I_\gamma\|,$$
$$\theta = \arctan(\|I_\alpha + I_\beta\|, I_\gamma), \text{ and } x = \Delta I_\alpha + \Delta I_\beta + \Delta I_\gamma,$$

$$\vec{F} = \vec{I} \cdot \vec{\alpha} \tag{58}$$
$$\Rightarrow \left[-\hat{r}(F_r + L_S\theta + L_x\varphi^2) + \hat{\theta}\left(F_s - L_x\varphi^2 \frac{r_2}{r}\right) + \hat{\varphi}\left(2F_s\varphi \frac{r_2}{r} + F_x\varphi\right)\right]$$

$$\vec{\nabla}\vec{F} = F_{\hat{r}} + \varphi\left(2F_S \frac{r_2}{r} + F_x\right) \tag{59}$$

$$\vec{\nabla} \cdot \vec{F} = \frac{1}{r}(1.5F_r - L_S - L_x\varphi^2)2\hat{r} + \frac{1}{x}\left(F_S \frac{r_2}{r} + 0.5F_x\right)2\hat{\varphi} \tag{60}$$

$$\vec{\nabla} \times \vec{F} = \frac{\hat{r}}{x}\left[\hat{\varphi}\left(2F_S\varphi \frac{r_2}{r} + F_x\varphi\right) + \hat{\theta}(2L_x\varphi) \frac{r_2}{r}\right] + \hat{\theta}\left[-\hat{r}(2L_x\varphi)\frac{1}{x} + \hat{\theta}(3F_S)\frac{1}{r}\right] + \frac{\hat{\varphi}}{r}\left[\hat{\theta}(F_S) + \hat{r}(L_S)\right] \tag{61}$$

ACKNOWLEDGMENT

This work used the Extreme Science and Engineering Discovery Environment (XSEDE), which is supported by National Science Foundation grant number ACI-1548562. The authors acknowledge the Texas Advanced Computing Center (TACC) at The University of Texas at Austin for providing HPC resources that have contributed to the research results reported within this paper. URL: http://www.tacc.utexas.edu